\documentclass[aip,pop,12pt,reprint]{revtex4-1}
\usepackage[dvips]{graphicx}
\usepackage{pgfgantt}
 \usepackage{delarray}
\usepackage{amsmath}
\usepackage{amssymb}
\newcommand{\g}[1]{\mbox{\boldmath $#1$}}

\usepackage{ae}
\usepackage{color}
\usepackage{ifthen}
\usepackage{bm}
\usepackage{ifpdf}
\ifpdf
	\usepackage[pdftex]{hyperref}
	\hypersetup{colorlinks=true,
		pdfstartview=FitV,
		linkcolor=blue,
		citecolor=red,
		urlcolor=blue,
	}
	\DeclareGraphicsExtensions{.pdf}
\else
	\usepackage[dvips]{graphicx}
	\usepackage{epsfig}
	\DeclareGraphicsExtensions{.eps}
	\usepackage{url}
\fi

\newcommand{\beq}{\begin{equation}}
\newcommand{\eeq}{\end{equation}}

\newcommand{\chalmers}{Department of Applied Physics, Chalmers University of Technology, Gothenburg, Sweden}
\usepackage{units}
\begin{document}
\title{Vlasov modelling of laser-driven collisionless shock acceleration of protons}
\author{B. Svedung Wettervik} \affiliation{\chalmers}
\author{T. C. DuBois} \affiliation{\chalmers} \author{T. F\"ul\"op}
\affiliation{\chalmers} \date{\today}
\begin{abstract}
  Ion acceleration due to the interaction between a short
  high-intensity laser pulse and a moderately overdense plasma target
  is studied using Eulerian Vlasov-Maxwell simulations.  The effects
  of variations in the plasma density profile and laser pulse
  parameters are investigated, and the interplay of collisionless
  shock and target normal sheath acceleration is analyzed.  It is
    shown that the use of a layered-target with a combination of light and
    heavy ions, on the front and rear side respectively, yields a
    strong quasi-static sheath-field on the rear side of the heavy-ion
    part of the target. This sheath-field increases the energy of the
    shock-accelerated ions while preserving their mono-energeticity.
\end{abstract}
\maketitle
 
\section{Introduction}
When multiterawatt laser {pulses} focused to ultrahigh intensities
illuminate the surfaces of dense plasma targets, protons can be
accelerated to energies of several tens of MeV within acceleration
distances of only a few micrometers \cite{daido,macchi}. There are
many potential applications for such beams, for example: isotope
generation for medical applications \cite{spencer}, ion therapy
\cite{orecchia,amaldi,bulanov,malka} and proton radiography
\cite{borghesi}. However, several of the foreseen applications of
laser-driven ion sources require high energies per nucleon (above 100
MeV) and a small energy spread, which is still far beyond the reach of
current laser-plasma accelerators.  It is therefore important to find
ways to optimize the acceleration process with the aim of producing
high-energy, mono-energetic ions.

At present, the most studied mechanism for laser-driven ion
acceleration is Target Normal Sheath Acceleration (TNSA) \cite{wilks},
which has been used to explain experimental results for laser
intensities in the range
$I=\unit[10^{18}\text{--}10^{20}]{W/cm}^2$. In TNSA, fast electrons
that are accelerated by a laser pulse set up an electrostatic
sheath-field that in turn accelerates ions from the rear side of the
target. Although the sheath-field is very strong (of the order of
teravolts/meter), the spatial extent and duration of the field is
short. Due to the short acceleration distance and time, it is
difficult to reach the high energies that are required for many
applications. Furthermore, TNSA yields protons with a broad energy
spectrum.  In contrast to this, electrostatic shock acceleration has
been suggested as a mechanism to obtain proton beams with a narrow
energy spectrum \cite{silva}. Experimental results have shown that
mono-energetic acceleration of protons can be achieved in
near-critical density plasma targets at modest laser intensities
\cite{haberberger}, with the hypothesis that these mono-energetic
beams are the result of shock-acceleration.

In hot and moderately overdense plasmas, shockwaves are of a
collisionless nature. The laser light pressure compresses the
laser-produced plasma and pushes its surface to high speed.  In the
electrostatic picture, ions are reflected by a moving potential
barrier and as long as the shock-velocity $v_s$ is constant, the
reflected ions obtain twice this velocity. The number of reflected
ions is dependent on the size of the potential barrier and temperature
of the ions. Macchi {\it et al.}  \cite{macchipre} reported that the
reflection of ions influences {the shock-wave},
yielding a trade-off between a mono-energetic spectrum and the number
of accelerated particles.  Additionally, Fiuza {\it et al.}
\cite{fiuza,fiuzaprl} have shown that if the sheath-field at the rear
side can be controlled, e.g. by keeping it approximately constant in
time by creating an exponentially decreasing density gradient at the
rear side, then the mono-energeticity of the ion distribution created
by reflection at the shock-front can be preserved.

%One of the drawbacks of collisionless shock acceleration (CSA) is that
%it normally results in low maximum proton energies at moderate laser
%intensities \cite{daido}. 
Combining collisionless shock acceleration (CSA) with a strong,
quasi-stationary sheath-field may be a way to {reach even higher maximum proton energies and} optimize the ion
spectrum.  In this work, we use 1D1P Eulerian Vlasov-Maxwell
simulations to study the interplay of CSA and TNSA. The objective is
to investigate how the efficiency of CSA is affected by variations in
the laser pulse and target parameters, and finding a way to
{tailor} the density profile of the target for enhanced ion
acceleration due to combined CSA and TNSA. It is shown that a layered
plasma target with a combination of light and heavy ions leads to a
strong quasi-static sheath-field, which induces an enhancement of the
energy of shock-wave accelerated ions.

The rest of the paper is organized as follows. In Section II we
describe the Vlasov-Maxwell solver {\sc Veritas} ({\bf V}lasov {\bf
  E}ule{\bf RI}an {\bf T}ool for {\bf A}cceleration {{\bf
    S}tudies}), used for modelling laser-based ion
acceleration. Section III presents results of simulations of the
interaction of short laser pulses with moderately overdense targets
with various density profiles. Section IV describes laser-driven ion
acceleration using multi-ion species layered targets.  Conclusions are
summarized in Section V.

\section{Numerical modelling}
Collisionless acceleration mechanisms can be modelled by the
Vlasov-Maxwell system of equations. Numerical approaches to solve this
system are primarily divided into Particle-In-Cell (PIC) methods and
methods that discretize the distribution function on a grid, so-called
Eulerian methods.  As PIC methods do not require a grid in momentum
space, they are efficient at handling the large range of scales
associated with relativistic laser-plasma interaction. They are
therefore very useful to model high dimensional problems. However, they
introduce statistical noise -- making it difficult to resolve the fine
structures of the distribution function.  On the other hand, solving
for the distribution function on a discretized grid yields a high
resolution of fine structures, but at a higher computational cost.  In
cases when the number of accelerated particles is low, as is sometimes
the case in collisionless shock acceleration (e.g.~in the experiment
described in Ref.~\citenum{haberberger}), the low-density tail of the
particle distribution is difficult to resolve in
PIC-simulations. Furthermore, the shock-dynamics may be affected by
the low-density non-thermal component in the ion distribution
\cite{macchipre}.  We therefore choose to implement the Eulerian
approach in this work.

For the case of a plasma with spatial variation in one direction, the
Vlasov-equation can be reduced to a two dimensional 1D1P problem:
\begin{equation}
\frac{\partial f}{\partial t}+\frac{p_x}{m\gamma}\frac{\partial f
}{\partial x}+q\left[E_x+\frac{1}{m\gamma}(\g p\times \g
  B)_x\right]\frac{\partial f }{\partial p_x}=0,
\end{equation}
where $f$ is the electron or ion distribution function, $x$ is a
spatial coordinate, $p_x$ is a momentum coordinate in this direction,
$m$ denotes the rest mass of the charged particles (electrons or ions) and
$\gamma$ is the relativistic factor.  The {single-particle}
Hamiltonian
$H=mc^2\left[1+(\g \Pi-q \g A)^2/m^2c^2\right]^{1/2}+q\phi$ yields
conservation relations for the transverse canonical momentum {(orthogonal to the direction of variation of the plasma)}:
$ \g \Pi_\perp=q\g A_\perp+\g p_\perp=0$. {The conservation of
$\g \Pi_\perp$ stems from the fact that the $y$ and $z$ coordinates do
not enter the Hamiltonian.} Here, $c$ is the speed of light, $q$ is the
charge, $\phi$ and $\g A$ are the electrostatic and vector potentials,
respectively.

The numerical tool used in this paper, {\sc Veritas}, employs
time-splitting~\cite{huot,knorr,sircombe,Ghizzo_2,Arber_1,filbet,sonnendrucker}
and the positive and flux conservative~\cite{filbet} methods to solve
the Vlasov-equation self-consistently with Maxwell's equations.  {\sc
  Veritas} has been extensively benchmarked by comparing with results
obtained by the PIC code {\sc Picador}\cite{picador} and results of
another Vlasov-Maxwell solver \cite{grassi}. Furthermore {\sc Veritas}
shows excellent agreement with analytical results derived in
Ref.~\citenum{Arkady}, where quasi-stationary solutions were obtained
for a cold overdense plasma with a fixed ion background, illuminated
with circularly polarized light {(the specifics of these benchmarks
  will be discussed in future work)}.

\paragraph*{The time-splitting method}
A common approach to solve the Vlasov-Poisson and Vlasov-Maxwell
systems is the time-splitting method.  Under this scheme the
Vlasov-Maxwell system is considered in the form:
\begin{equation}
\frac{\partial f}{\partial t}+\mathcal{L}f=0,\label{tss}
\end{equation}
where, in the 1D1P case:
\begin{equation}
\mathcal{L}=\frac{p_x}{m\gamma}\frac{\partial}{\partial x}+q\left[E_x+\frac{1}{m\gamma}(\g p\times  \g B)_x\right]\frac{\partial}{\partial p_x}.
\end{equation}
Writing $\mathcal{L}=A+B$, we introduce the two
equations:
\begin{equation}
\frac{\partial f}{\partial t}+Af=0 \label{tss1}
\end{equation}
and
\begin{equation}
\frac{\partial f}{\partial t}+Bf=0.\label{tss2}
\end{equation}
Equation~\eqref{tss} is advanced to second order accuracy in time by
first advancing Eq.~\eqref{tss1} a half time-step, followed by
advancing Eq.~\eqref{tss2} a full time-step and finally advancing
Eq.~\eqref{tss1} yet another half time-step. In addition to this, the
electromagnetic field is advanced and defined at half-integer
time-steps.

Time-splitting can be performed using different choices of the
operators $A$ and $B$. In this paper, we use
\begin{align}
\label{split1}
A&=\frac{p_x}{m\gamma}\frac{\partial}{\partial x}+\frac{\partial}{\partial x}\left(\frac{p_x}{m\gamma}\right),\nonumber\\
B&=q\left[E_x+\frac{1}{m\gamma}(\g p\times
  \g B)_x\right]\frac{\partial}{\partial p_x}\nonumber\\
  &+\frac{\partial}{\partial p_x}\left\{q\left[E_x+\frac{1}{m\gamma}(\g p\times
  \g B)_x\right]\right\}.
\end{align}
This yields  the split equations:
\begin{equation}
\frac{\partial f}{\partial t}+\frac{\partial}{\partial x}\left(\frac{p_x}{m\gamma}f\right)=0 \label{tss1a}
\end{equation}
and
\begin{equation}
\frac{\partial f}{\partial t}+\frac{\partial}{\partial p_x}\left\{q\left[E_x+\frac{1}{m\gamma}(\g p\times
  \g B)_x\right]f\right\}=0,\label{tss2b}
\end{equation}
which conserve particle number individually.  Further details are given in Appendix A.

\paragraph*{Electromagnetic fields}
For a one-dimensional system, Maxwell's equations take the form
\begin{equation*}
\frac{\partial B_x}{\partial x}=0,\quad \frac{\partial B_x}{\partial t}=0,
\end{equation*}
\begin{equation*}
\frac{\partial B_y}{\partial t}=\frac{\partial E_z}{\partial x},\quad \frac{\partial B_z}{\partial t}=-\frac{\partial E_y}{\partial x},
\end{equation*}
\begin{equation*}
\frac{\partial E_x}{\partial x}=\rho/\epsilon_0,\quad \epsilon_0\mu_0\frac{\partial E_y}{\partial t}=-\mu_0J_y-\frac{\partial B_z}{\partial x}
\end{equation*}
and
$$\epsilon_0\mu_0\frac{\partial E_z}{\partial t}=-\mu_0J_z+\frac{\partial B_y}{\partial x}.
$$
Here, the currents and  charge density are determined by the
distribution function{s}, according to
$$ \g J_\perp =\sum_s \frac{q_s}{m_s}\int \frac{\g p_{\perp s}}{
  \gamma_s}f_s\,\text{d}p_x
$$
and
$$
\rho=\sum_s q_s \int  f_s \,\text{d}p_x,
$$ where the summation ranges over all species. The transverse  vector potential $\g A_\perp$ is obtained by
$ \g E_\perp=-\partial \g A_\perp/\partial t$ and is used together
with the conservation of canonical momentum $\g \Pi_\perp$ to
calculate the relativistic factor $\gamma$ and the transverse
components of the current. The numerical scheme for solving the
electromagnetic field equations is described in Appendix B.

\section{TNSA and shock-wave acceleration}
We consider moderately overdense plasma targets with different density
profiles (rectangular, exponential and multi-species layered) having
peak { number} densities $n_0=2.5 n_{c}$, where $n_{c}=m_e
\omega^2\epsilon_0/e^2$ is the {\it cutoff} or {\it critical density}
at which the laser frequency $\omega$ equals the electron plasma
frequency. Ions are assumed to be cold, with an initial temperature
$T_i=\unit[1]{eV}$, {while} electrons are assumed to have an initial
temperature $T_e=\unit[5]{keV}$.  The targets are heated by linearly
polarized Gaussian laser-pulses with short pulse lengths, having full
width at half maximum (FWHM) {of the intensity} in the range $25-50$
fs. The Gaussian shape factor of the vector potential is
$a(t)=a_0\exp{[-2 \ln{2} (\tau/t_p)^2]}$, where $\tau=t- t_p$ and
$t_p$ is the pulse duration at FWHM.  The dimensionless laser
amplitude $a_0=e A_0/m_ec$ is in the range of $a_0=2.5-3.5$ and
relates to the laser intensity $I$ and wavelength $\lambda$ according
to $a_0=0.85 (I \lambda^2/10^{18} \mbox{W} \mbox{cm}^{-2}\mu
\mbox{m}^2)^{1/2}$.  The combination of $a_0$ and pulse length is
varied such that the laser fluency $\mathcal{F}=T^{-1}\int
a(t)^2\,\text{d}t$ remains constant. Here, $T$ is the {duration of
  the optical cycle} corresponding to the wavelength
$\lambda$. {Regarding numerical resolution, simulations
  have been performed with spatial resolution $\Delta x=\lambda/200$,
  momentum space resolution $\Delta p=m_ec/20$ and time step $\Delta
  t=T/200$. }

\subsection{Density profile variation}
The target is assumed to be a proton-electron plasma, i.e. with $Z=A=1$,
where $Z$ and $A$ are the charge and mass numbers, respectively.
The  plasma density profile is taken to be
\begin{equation}
 n(x) = \left\{\def\arraystretch{1.2}%
  \begin{array}{@{}c@{\quad}l@{}}
    n_0(x-2\lambda)/\lambda & \text{if } x\in [2\lambda,3\lambda]\\
    n_0\exp\left[-(x-3\lambda)/5\lambda\right] & \text{if } x\in [3\lambda,25\lambda]\\
  \end{array}\right.
\label{haberberger}
\end{equation}
with $n_0=2.5n_c$ {for both electrons and ions}. This type of density
profile, with a linear rise on the front side and an exponential
decrease on the rear side, can be naturally formed by the pre-heating
and expansion of the target due to a laser pre-pulse
\cite{haberberger}.

We use a linearly polarized laser pulse with $a_0=2.5$ and
pulse-length of \unit[50]{fs}. For reference, the amplitude peak of
the laser-pulse impinges on the front side of the plasma at time
$t=t_p$. The wavelength is taken to be
  $\lambda=\unit[0.8]{}\mu$m.

\begin{figure}[htbp]
   \includegraphics[width=\columnwidth]{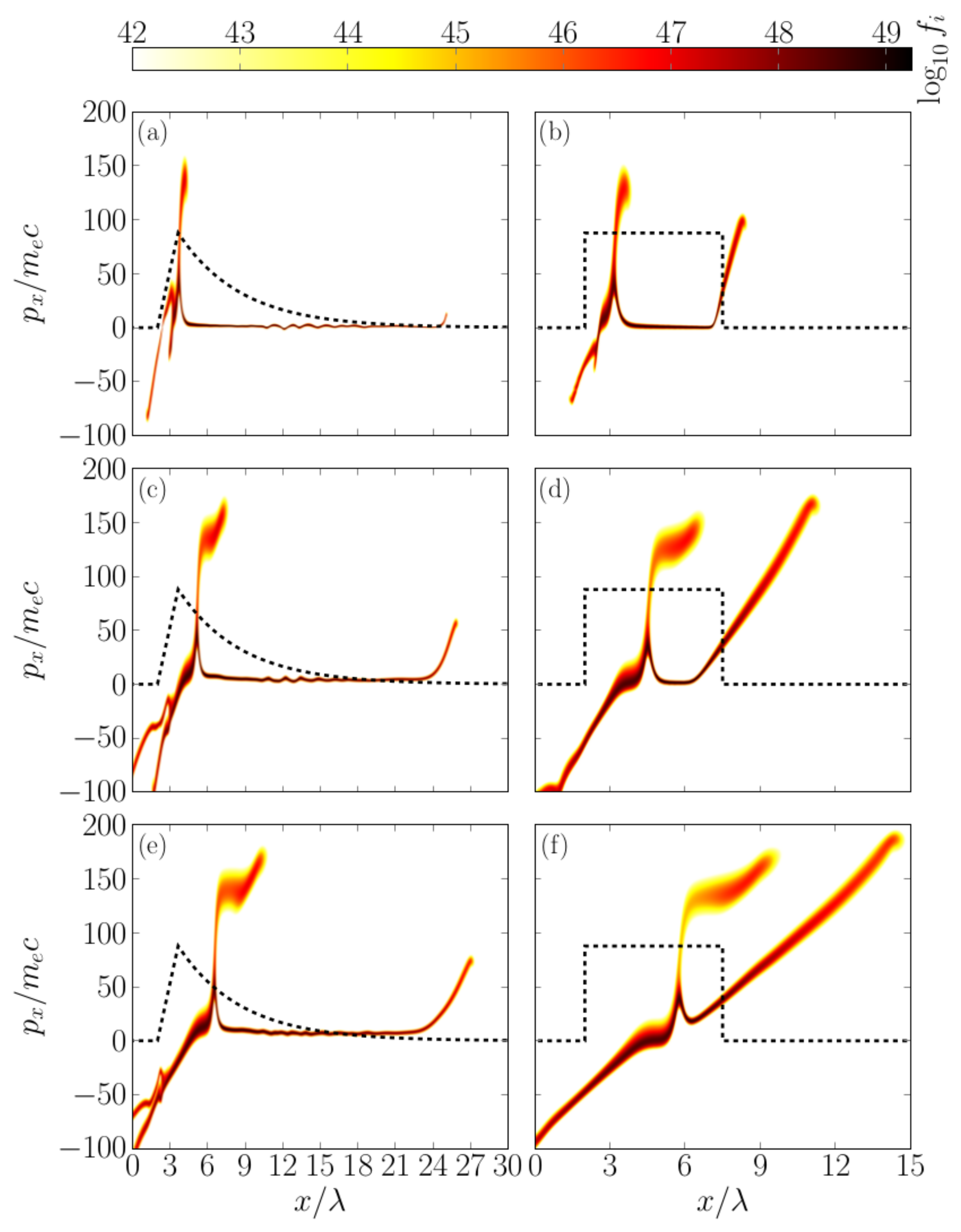}
\caption{\label{ionsandelectrons}Ion phase-space distribution at three
  different time instants ({ $t=39T$, $75T$
and $108T$}) for the exponential (left) and
  rectangular (right) plasma density profiles. The target is
  irradiated by a linearly polarized pulse with $a_0=2.5$ and pulse
  length $50$ fs. }
\end{figure}

  At incidence of the laser-pulse on the target, the laser energy is
  absorbed near the critical density and electrons are accelerated to
  strongly relativistic energies. The left panel of
  Fig.~\ref{ionsandelectrons} shows the ion phase-space distribution
  $f_{i}(x,p_x)$ at the three time instances $t=39T$, $75T$ and
  $108T$.  The target is heated and an electrostatic shock-structure
  is generated that travels into the plasma at a constant velocity
  $v_s=0.041c$. The velocity of the shock-wave is {inferred} from
  the velocity of the maximum of the electrostatic potential
  barrier. This value can be compared to the hole-boring velocity
  $v_{\rm HB}=0.034 c$, obtained via \cite{macchipre}
  {    $v_{\rm HB}=a_0c
    \left[(Z/A)(m_e/m_p)(n_c/n_e)(1+R)/2\right]^{1/2}$,}
    with $n_e=2.5n_c$, $a_0=2.5$, $Z=1$ and
    $m_p/m_e=1836$. { From simulations we determined the reflectivity to be $R=0.67$. } The reflected ions
    initially travel with { a momentum corresponding to} twice the
    shock-velocity $p\simeq 130 m_e c$, see
    Fig.~\ref{ionsandelectrons}a; however as time goes by, the ion
    spectrum becomes broader as shown in
    Fig.~\ref{ionsandelectrons}c\&e. The broadening of the ion
    spectrum is due to two different effects: First, not all the ions
    will have the same initial reflection velocity, because the speed
    of the potential barrier varies during its formation. Second, the
    reflected ions will be affected by the longitudinal electric
    field, which also varies {in both space and time}.  {At
      the rear side of the target, one observes TNSA, but the sheath
      field is not strong enough for substantial acceleration in this
      case.}

\begin{figure}[htbp]
  \includegraphics[width=\columnwidth]{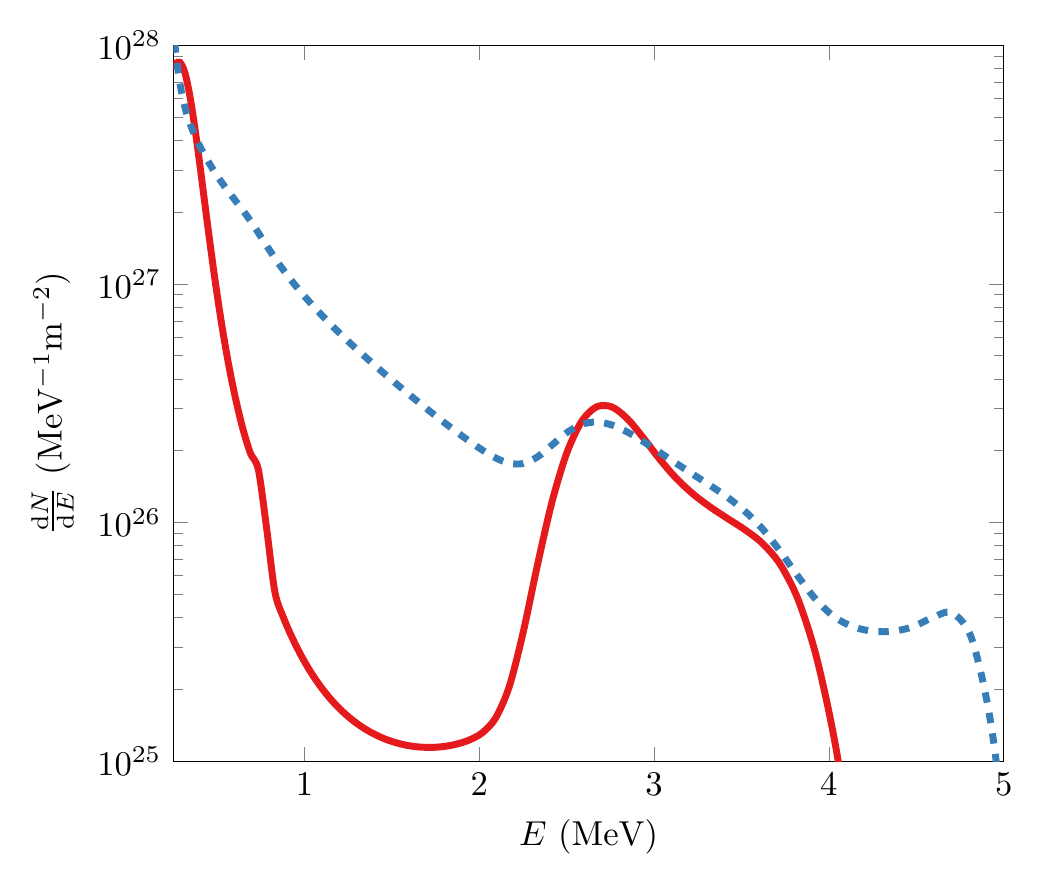}
\caption{\label{spectrum} Proton spectrum at $t=108T$ for the
  exponentially decreasing density profile (red solid) and the
  rectangular plasma slab (blue dashed). }
\end{figure}

To investigate the effect of the density profile, we also consider a
rectangular plasma slab with $n_0=2.5n_c$ and thickness
$d=5.5\lambda$. The length of the slab was chosen so that the particle
number is the same for both the rectangular and exponential density
profiles. The ion distribution at three time instances ($t=39T$, $75T$
and $108T$) is shown in the right panels of
Fig.~\ref{ionsandelectrons}. The dynamics of the shock-formation is
similar for both cases, but the shock-velocity is slightly lower
{and closer to the hole-boring velocity}. It is $0.037c$ in the
rectangular case compared with $0.041c$ for the exponential
case. Furthermore, the TNSA is stronger than in the exponential case,
resulting in a TNSA-dominated broad ion energy spectrum, {as can be seen in Fig.~\ref{spectrum} (blue dashed line)}. From this we
can conclude that the shape of the density profile at the rear side
has an important role in suppressing the sheath-field responsible for
TNSA. Similar conclusions were drawn in
Refs.~\citenum{fiuza,fiuzaprl}, where electrostatic shocks driven by
the interaction of two plasmas with different density and relative
drift velocity were studied using PIC simulations.

The energy spectrum given in Fig.~\ref{spectrum} was calculated based
on the entire ion population using
$\text{d}N/\text{d} E=\left(\text{d}p/\text{d}E\right)\int f_i
(x,p_x)\,\text{d}x$,
where $E=\sqrt{m^2c^4+p^2c^2}$ and $f_i (x,p_x)$ is the ion
distribution function. Note, that the exact value of the initial ion
temperature does not influence the results, as long as the ions are
cold at the start of the simulation. A simulation with the same
laser-pulse and target parameters, but an initial ion temperature of
$T_i=100\;\rm eV$ gives identical results.  This also applies to a
simulation with an initial electron temperature of $T_e=2.5\;\rm keV$.

\begin{figure}[htbp]
  \includegraphics[width=\columnwidth]{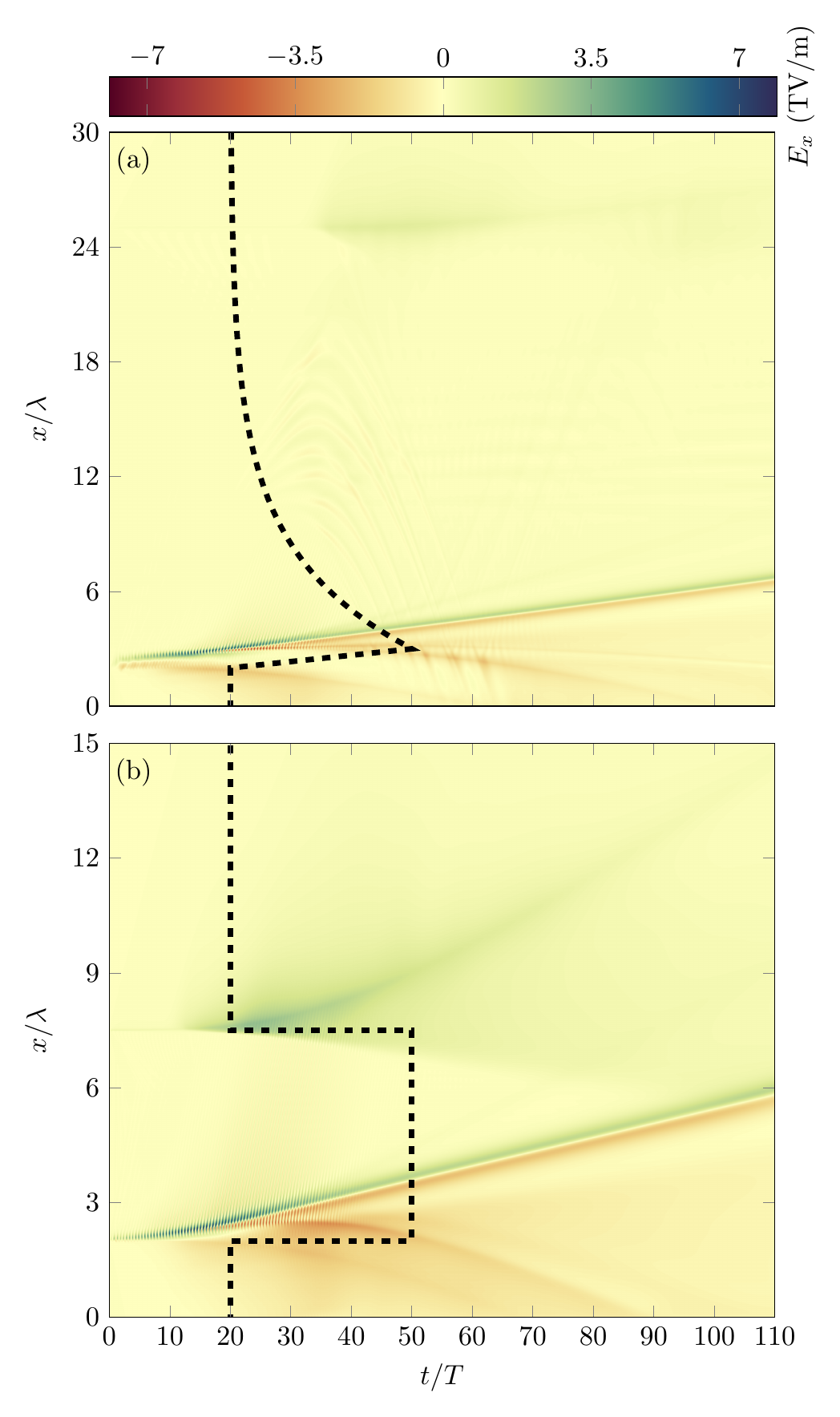}
\caption{\label{field} Longitudinal electric field as a function of
  position and time for the (a) exponential and (b) rectangular
   density profiles. The target is irradiated by a
  linearly polarized pulse with $a_0=2.5$ and pulse length
  $50$ fs.}
\end{figure}

Figure~\ref{field} shows the longitudinal electric field as a function
of time and space for the exponential and rectangular density
profiles. From Fig.~\ref{field}a it can be noted that the sheath-field
at the rear side of the exponential density profile is smaller than the one at the shock-front. This,
combined with the fact that regions with significant electric fields
at the rear side are associated with lower ion density, leads to less
pronounced TNSA. Therefore the resulting ion spectrum has a broad
bump-like structure with a maximum ion energy at around 3 MeV, as
shown in Fig.~\ref{spectrum} (red solid line); contrasted with the
rectangular density profile, for which the proton spectrum is shown
with a blue dashed line. In the latter case, the sheath-field at the
rear side is very strong, as can be seen in Fig.~\ref{field}b, and
gives rise to the broad exponential proton spectrum that is typical
for cases when TNSA is dominant.

\subsection{Laser pulse variation}

\begin{figure}[htbp]
 \includegraphics[width=\columnwidth]{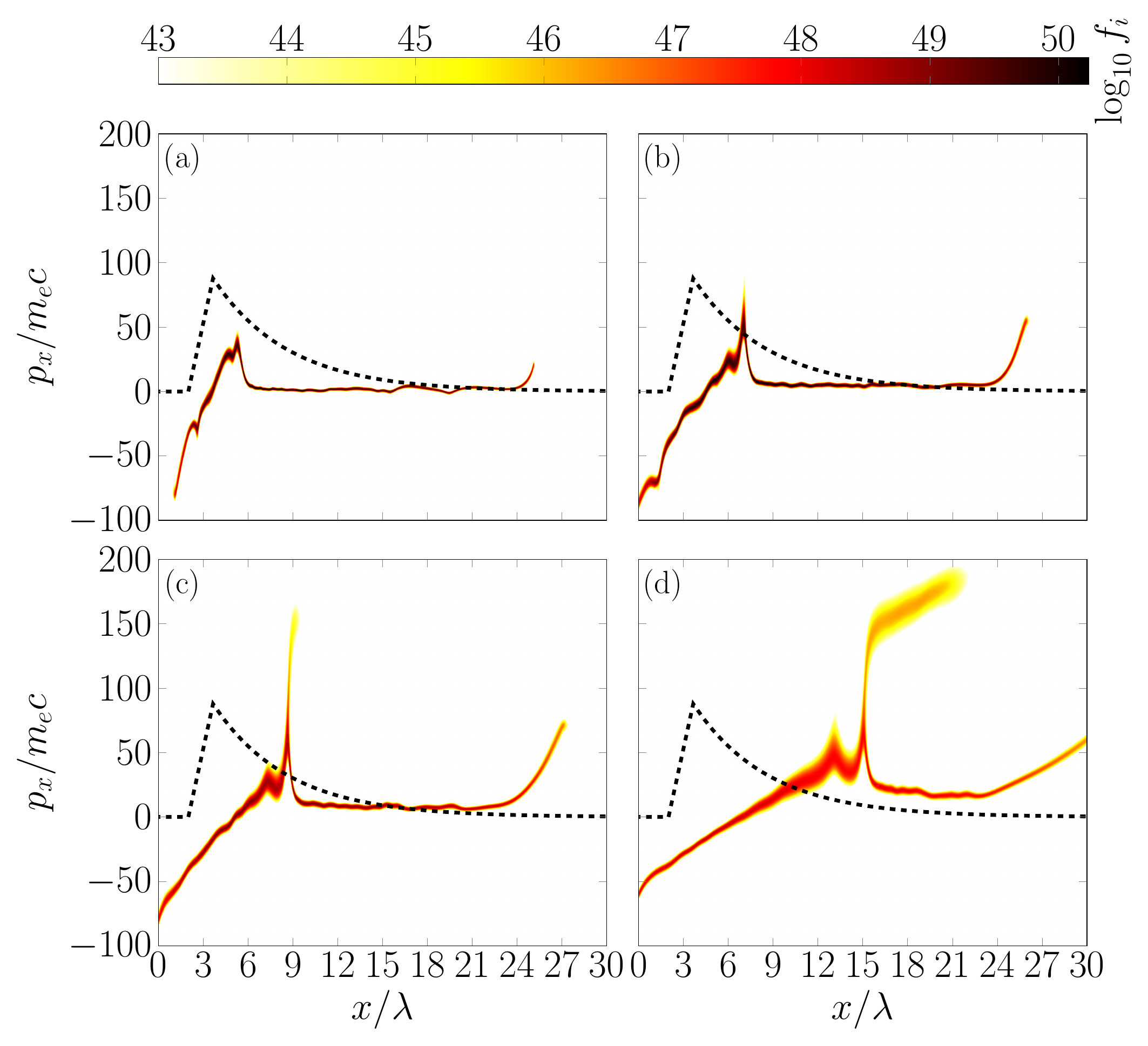}
\caption{\label{haberberg_intense} Ion phase-space distribution for
  the exponential density profile irradiated by a linearly 
  polarized pulse with $a_0= 2.5\sqrt{2}$ and pulse length
  \unit[25]{fs} {at  $t=39T$, $75T$, $108T$ and $240T$}.}
\end{figure}

\paragraph*{Pulse intensity} Here, we show results for the exponential
density profile heated by a laser pulse with $a_0=2.5\sqrt{2}$ and
pulse length  $\unit[25]{fs}$. With these parameters the laser
fluency is the same as in the case with the longer and less intense
pulse described in the previous subsection. Figure
\ref{haberberg_intense} shows the ion distribution function at times
$t=39T$, $75T$, $108T$ and $240T$. Compared to the case with the
longer pulse with lower intensity (both the rectangular and
exponential profiles), the time for the shock to develop is
considerably longer.  The velocity of the shock-wave in the more
intense pulse case is  also higher, with $v_s=0.049c$.

{At first it may seem counterintuitive that the shock is developed
  later in the case of the shorter and more intense pulse, given the
  fact that the shock-velocity is higher. The main reason for the
  later development is the shorter pulse length { and
    higher intensity}, which { leads to operation close to
    the onset of relativistic transparency and larger peneteration of
    the pulse into the plasma rather than reflection/compression at
    the plasma vacuum interface. This} gives smaller peak ion and
  electron densities after interaction with the laser pulse
  {and results in} differences in the electrostatic
  potential and the reflection time of the ions.}

\begin{figure}[htbp]
  \includegraphics[width=\columnwidth]{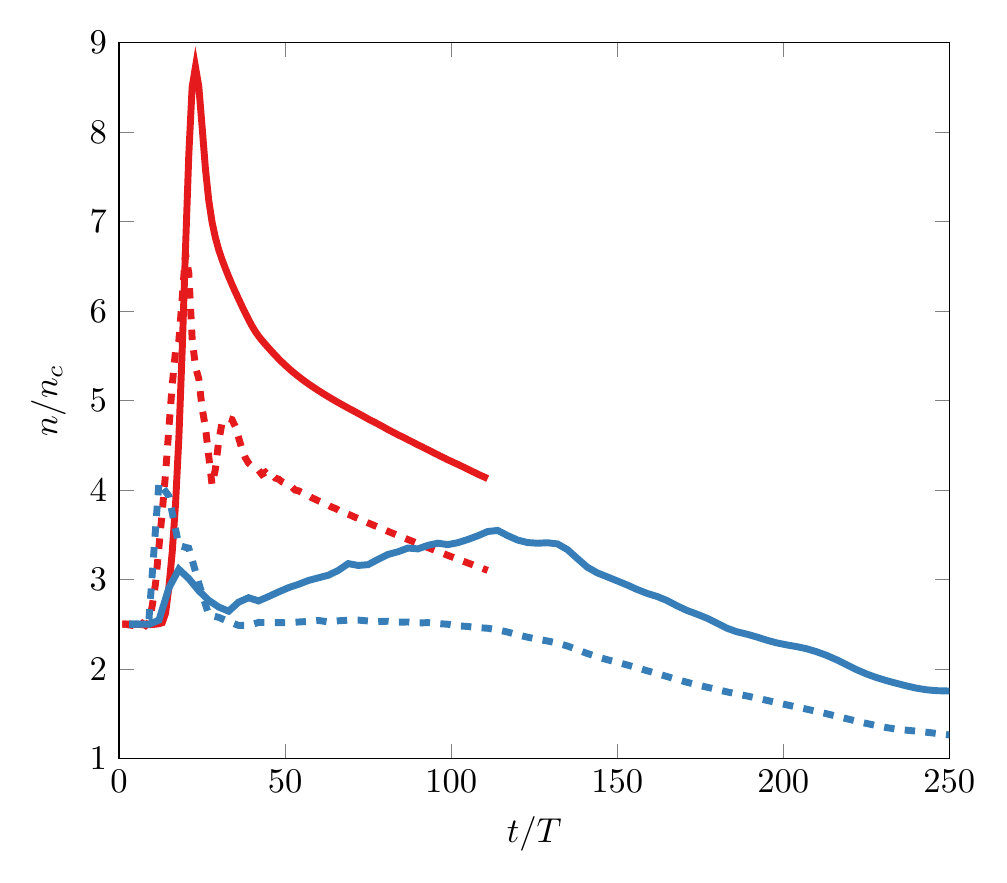}
  \caption{\label{peakdens} Peak ion (solid) and electron (dashed)
    densities as functions of time. Red lines: $a_0= 2.5$. Blue lines:
    $a_0= 2.5\sqrt{2}$.  }
\end{figure}

{In both cases the ion and electron densities have
  their largest peak value right after the interaction with the
  pulse. For the shorter pulse case however, the peak values are much
  lower (see Fig.~\ref{peakdens}). These lower ion and electron densities lead to a more gradual and wider potential barrier. }

\begin{figure}[htbp]
   \includegraphics[width=\columnwidth]{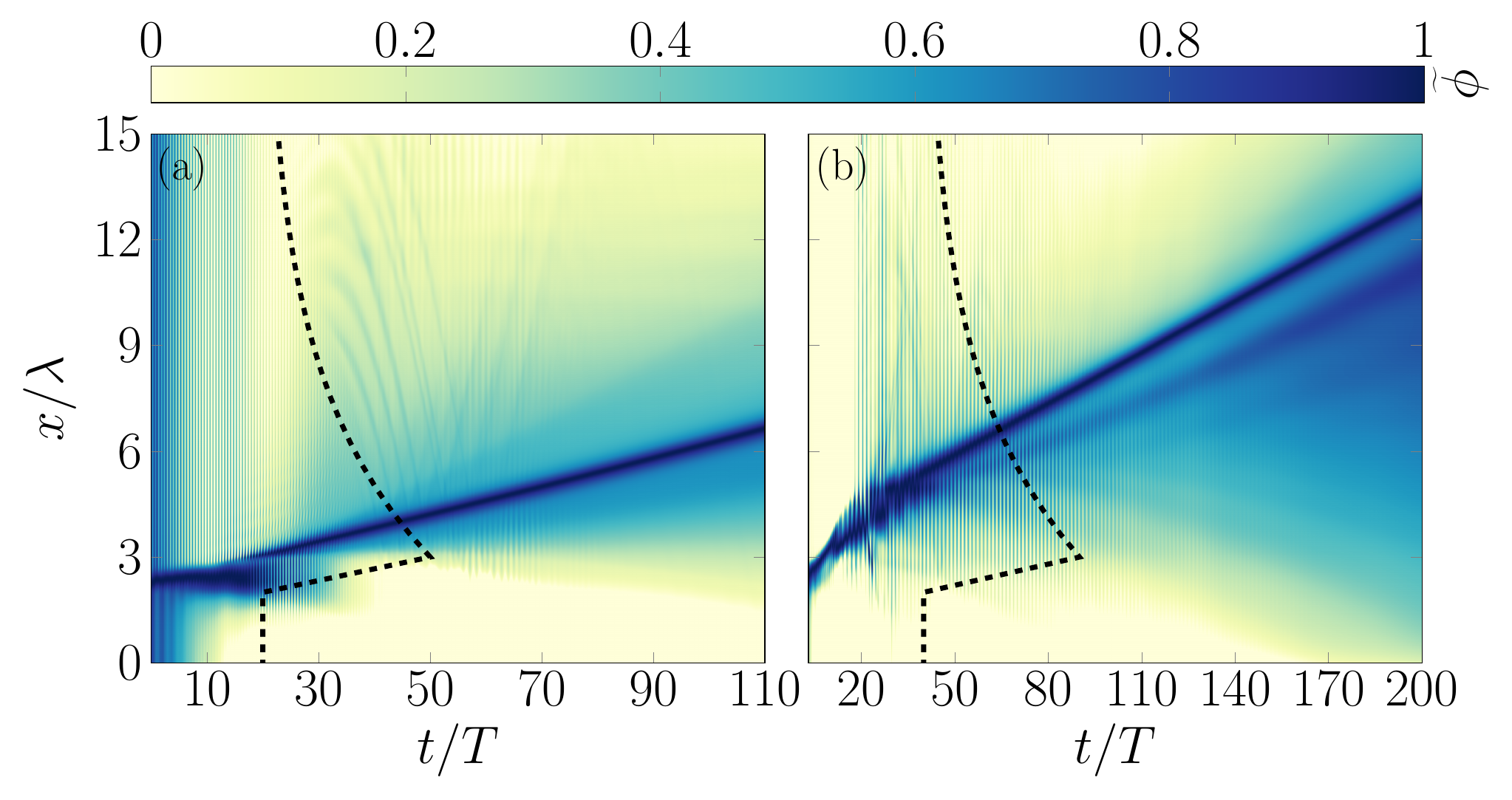}
\caption{Electrostatic potential as function of space and time. (a)
  $a_0= 2.5$ and pulse length \unit[50]{fs}, (b) $a_0= 2.5\sqrt{2}$
  and pulse length \unit[25]{fs}.}\label{potentials}
\end{figure}
Figure~\ref{potentials} shows the electrostatic potential as a
function of time and space for the exponential density profile in the
cases with different laser intensities and pulse lengths ($a_0=2.5$
left panel, $a_0=2.5\sqrt{2}$ right panel).  The potential is scaled
and shifted according to
$2e\phi/m_pv_s^2-\max{[2e\phi/m_pv_s^2]} + 1$, such that the peak of
the potential is unity.  Furthermore, the potential has been truncated
at zero.   When the laser pulse hits the target there is an oscillation in the
densities and electrostatic potential, due to
$\g j\times \g B$-heating. The frequency of the oscillation is
approximately twice the laser frequency.  For the more intense pulse
these oscillations persist for a longer time.

For ions at a low temperature, reflection occurs for a potential
barrier height approximately equal to unity.  If the ions have
acquired a velocity in the direction of the shock, reflection occurs
at a slightly smaller barrier height, which is the case at later times
due to sheath-expansion of the target.  Hence, a shock solution may
develop although the potential barrier initially does not have
sufficient height for ion reflection to occur.

{For a linearly rising potential barrier, the
  reflection time for an ion is given by $d/v_s$, where $d$ is the
  spatial extent of the barrier. As mentioned before, the potential
  barrier for the shorter and more intense pulse is initially wider
  (see Fig.~\ref{potentials}), yielding a proportionally longer
  reflection time.  Even if the shock-velocity is slightly higher in
  the more intense case, the spatial extent of the barrier is even
  larger, so the reflection time, $d/v_s$, is longer.  }

{ Figure~\ref{peakdens} shows a steepening of the peak ion
  density from $t \simeq 25 T$ to $t\simeq 100 T$ in the case of the
  shorter and more intense pulse (solid blue line).  This steepening
  is associated with persisting oscillations in the electrostatic
  potential. The width of the potential barrier is reduced and therefore the reflection time for the ions as well.  Finally, a shock is
  developed, albeit much later than in the case of the longer pulse. {This shows that shock acceleration can be operated close to the relativistic transparency  regime which maximizes the hole-boring velocity and is also seen to yield a higher shock velocity.}}

\paragraph*{Pulse splitting}
Previous numerical results indicate that using a train of short laser
pulses may produce more efficient ion-acceleration than one Gaussian
pulse with the same energy \cite{mironov,yu,markey}. Furthermore,
experimental results in Ref.~\citenum{haberberger} show that a smooth
pulse containing the same energy as a pulse train will result in a
monotonically decreasing ion spectrum, instead of a spectrum with a well-defined
peak as in the pulse-train case. This  indicates that the efficiency
of shock-acceleration is improved in the case of multiple pulses.  

\begin{figure}[htbp]
   \includegraphics[width=\columnwidth]{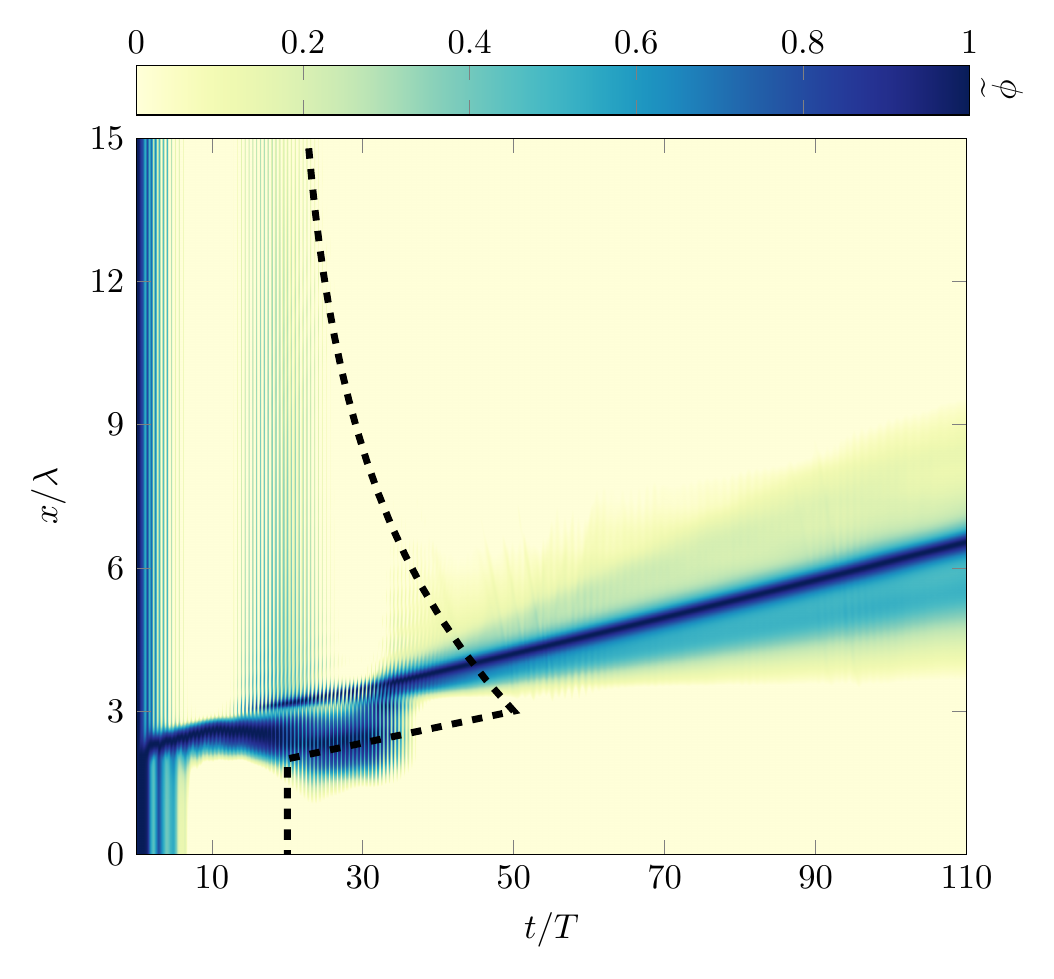}
\caption{\label{split} Electrostatic potential in the case
  of the exponential density profile irradiated by two laser pulses
  with $a_0=2.5$ and pulse lengths $\unit[25]{fs}$, that are
  separated by $\unit[50]{fs}$ in time.}
\end{figure}

To investigate how the splitting of the pulse affects the
shock-dynamics, here we consider the exponential density profile
irradiated by two laser pulses with $a_0=2.5$ and pulse length
$\unit[25]{fs}$, that are separated by $\unit[50]{fs}$ in time.
Figure \ref{split} shows the electrostatic potential as a
function of time and space.  The variation in the electrostatic
potential indicates that a shock-structure is formed already after the
first pulse. The arrival of the second pulse perturbs the potential
barrier associated with the shock, leading to a slight increase of its
velocity, from $v_s\approx 0.030c$ to $v_s\approx 0.039c$.  Hence, the
use of two pulses increases the velocity of the shock, although it remains smaller than if all the
energy would have been in a single pulse, {c.f. $v_s \approx 0.041c$
  for a single pulse with $a_0=2.5$ and pulse length 50 fs.}

The effect of pulse-splitting is even more important for higher
initial plasma densities, as predicted by previous numerical and
experimental results, see e.g. Ref.~\citenum{markey}. The reason is
that the absorption of the second pulse can be enhanced {if the
  target density at the front side becomes lower due to
  heating-induced expansion caused by the first pulse}. This results
in higher electron temperatures and consequently stronger TNSA. This
heating-induced absorption enhancement effect is not as pronounced if
the initial densities are close to the critical density. Our
simulations show that for a rectangular density profile with $n_0=25
n_c$, the energy spectrum is TNSA-dominated and the cutoff ion energy is increased by 10\% in the case of two pulses with $a_0=2.5$ and pulse lengths 25 fs separated in time
with 50 fs, compared with the case of one pulse with $a_0=2.5$ and
pulse length 50 fs. Corresponding simulations with peak densities
$n_0<2.5 n_c$ do not give a substantial increase in the proton energy
if the pulse is split.

\begin{figure}[htbp]
      \includegraphics[width=\columnwidth]{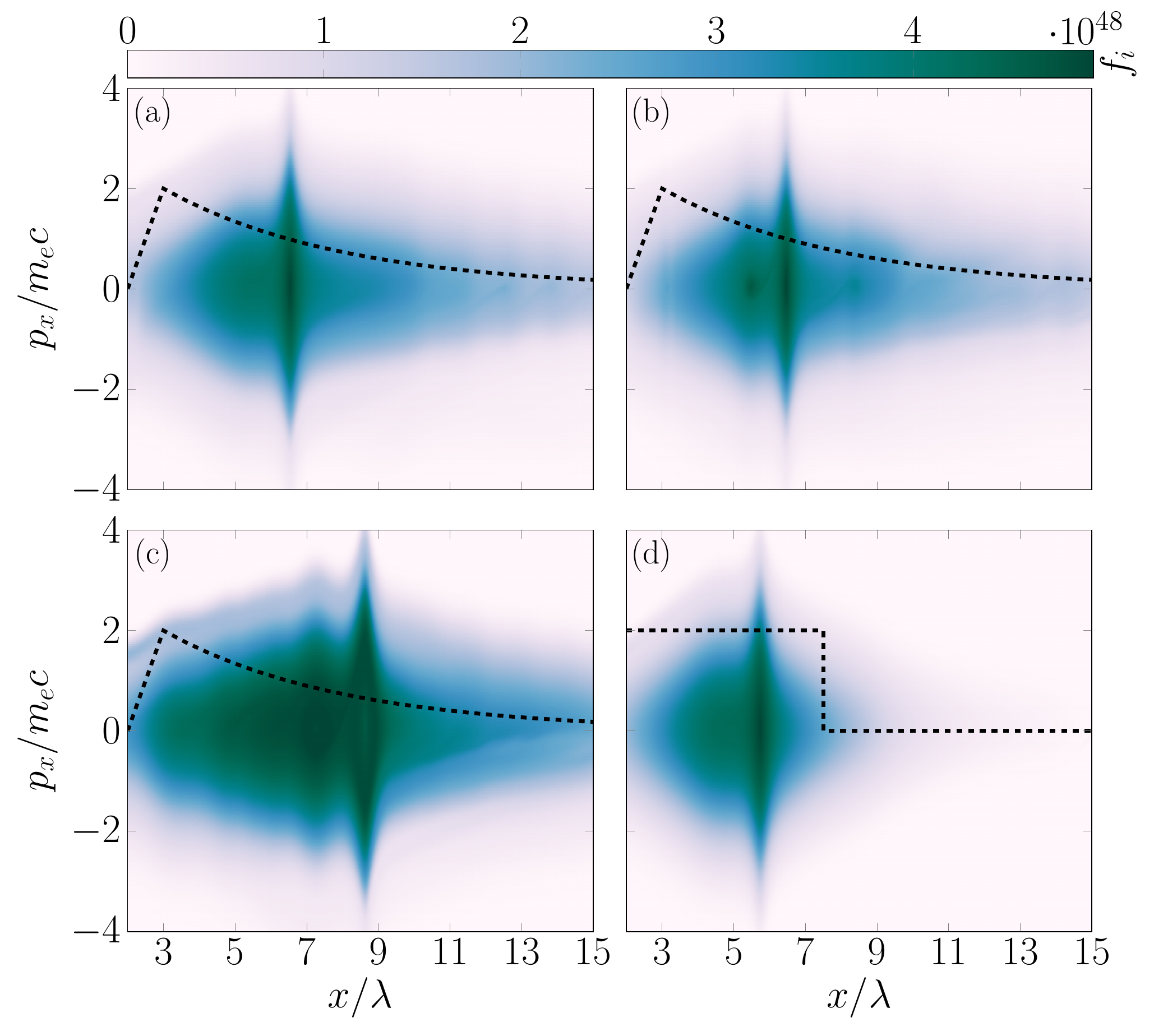}
      \caption{\label{electrons} Electron phase-space distribution {at  $t=108T$} in
        the exponential and rectangular density profile cases. Panels
        (a) and (d) show the case with $a_0=2.5$ and pulse length
        $\unit[50]{fs}$, for the exponential (a) and rectangular (d)
        plasma profiles. Panel (b) depicts the exponential density
        profile irradiated by two laser pulses with $a_0=2.5$ and
        pulse length $\unit[25]{fs}$, that are separated by
        $\unit[50]{fs}$ in time. Panel (c) is for the exponential
        profile with $a_0= 2.5\sqrt{2}$ and pulse length
        \unit[25]{fs}.}
\end{figure}

Figure~\ref{electrons} shows snapshots at $t=108T$ of the electron
distribution function for both the exponential and the rectangular
profiles for different values of $a_0$ and pulse shapes. The
simulations confirm that the hot electron temperature in all cases is
on the order of magnitude of the ponderomotive scaling
$T_{h}\approx m_e c^2(\sqrt{1+a_0^2/2}-1)$. Specifically, the case with the
more intense pulse ({with shortest } pulse length), leads to {the
  highest} hot electron temperature, as expected from the
ponderomotive scaling, even if the fluency is the same in the {
  different} cases.  The Mach number of the shocks $M=v_s/c_s$ is
around 1.7 in all cases, if we use the ponderomotive scaling to
estimate the hot electron energy as the temperature in the sound speed
$c_s=\sqrt{T_e/m_i}$.

\section{Enhanced ion acceleration using multi-ion species layered targets}
\begin{figure*}[htbp]
 \includegraphics[width=\textwidth]{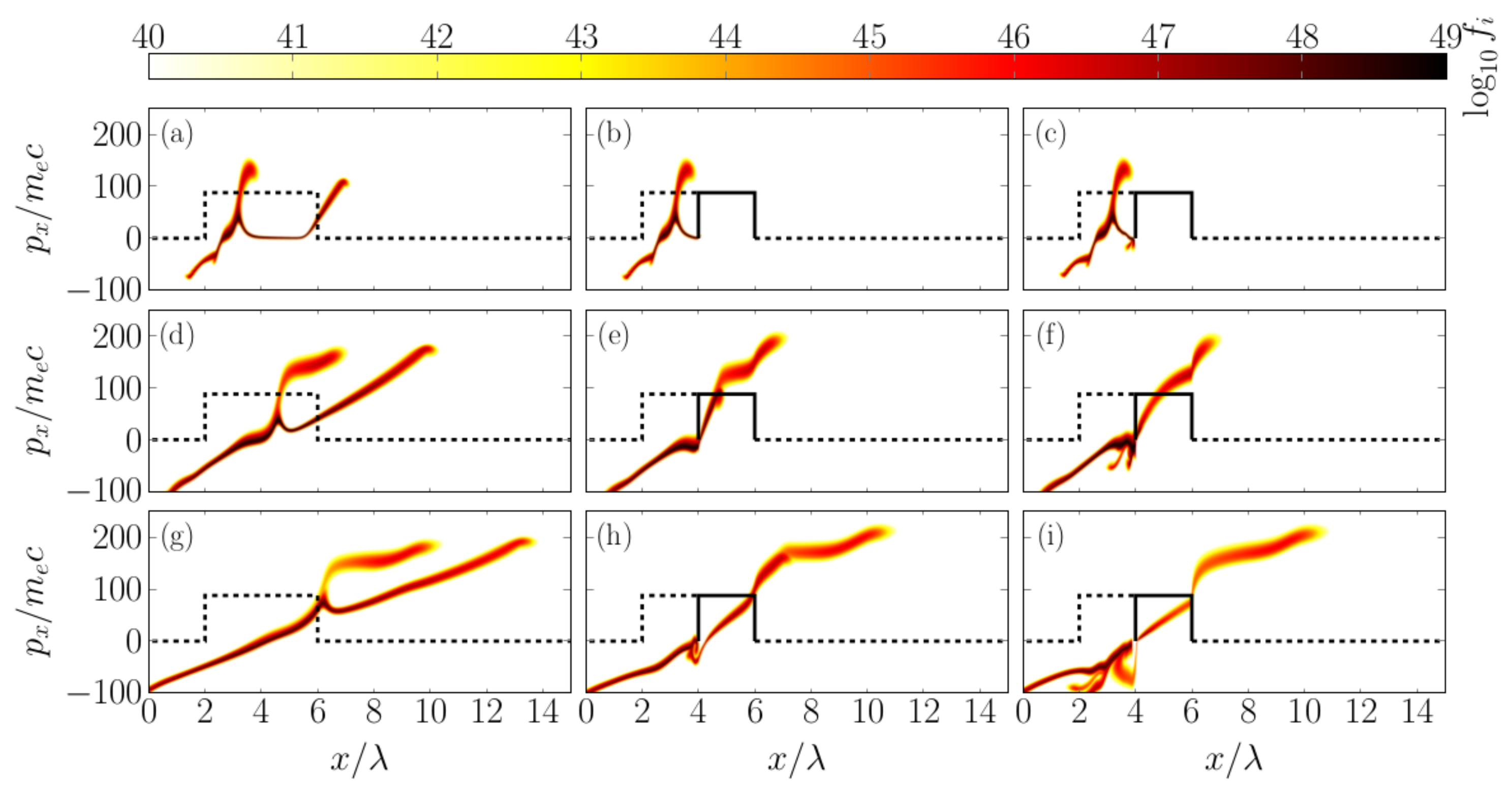}
 \caption{\label{layers} Ion phase-space distribution for single- and
   double-layer target structures, irradiated by a linearly polarized
   pulse with $a_0= 2.5$ and pulse length \unit[50]{fs} {     at $t=39T$, $75T$ and $108T$}. (a,d,g) show the single-species
   target, (b,e,h) the layered target with $n=2.5n_c$ and (c,f,i) the
   layered target with the high density heavy ion layer having
   $n=25n_c$. }
\end{figure*}
For a target with a steep rear boundary, a strong sheath-field can be
obtained and used to increase the energy of shock-accelerated
ions. Targets with a single light ion species are subject to
significant TNSA and hence the resulting ion energy-spectrum becomes
broad. Furthermore, the acceleration of ions at the rear side leads to
a decay of the sheath-field strength and hence its usefulness for
post-acceleration is reduced.  To combine the use of a strong
sheath-field for post-acceleration and a low degree of TNSA, we
consider a double layered rectangular target consisting of a layer of
light ions (protons) at the front side ($x\in [2\lambda,4\lambda]$)
and heavy (immobile) ions at the rear side ($x\in
[4\lambda,6\lambda]$). We use the density profile $n(x)=2.5n_c$ in the
light ion part. In the heavy ion part we consider two cases for the
electron density profile, $n(x)=2.5n_c$ and $n(x)=25n_c$,
respectively.  For comparison, we also consider a single layer
rectangular target with protons $n(x)=n_0$ for $x\in
[2\lambda,6\lambda]$. The targets are irradiated by a laser pulse with
$a_0= 2.5$ and pulse length $\unit[50]{fs}$.

Figure~\ref{layers} shows snapshots of the ion distribution function
for the single-species and double layered targets at $t=39T$, $75T$
and $108T$. The laser heats the front side of the target and launches
a shock. Until the shock reaches the region with heavy ions in the
double layer, its behaviour is similar to that in the single species
target.  For the double-layered targets the shock wave is stopped at
the interface between the layers, but the shock-wave reflected ions
continue and finally cross the rear side of the target. When this
occurs the ions are further accelerated due to the sheath field,
leading to higher proton energies than what they would have from the
reflection by the shock-wave alone.

If the heavy ion layer has higher density than the light ion layer,
ions can be slowed down due to the sheath field that is created by the
density difference at the interface. Those ions that have acquired
enough energy from the shock-wave potential barrier can penetrate the
interface and continue through the target. { The interface
  between the layers acts effectively as a filter: it reflects the low
  energy ions, and leads to a narrower energy spectrum after the
  interface.} By comparing Figs.~\ref{layers}(h,i) we see that more
protons penetrate the interface in the low density case{,
  as can be expected since the size of the potential barrier
  associated with the sheath field at the interface between the light
  ion and heavy ion layers is smaller in this case.

  Inside the heavy ion layer, the energy spectrum ranges from zero for
  protons that had initial energy just above the threshold for
  reflection, to the highest energy of reflected ions, reduced by the
  size of the potential barrier. The electric field inside the heavy
  ion layer is very small, so the protons are crossing this layer
  without gaining much energy. As it takes less time for the higher
  energy light ions to cross the heavy ion layer, the distribution is
  rotated in phase-space, as can be noted by comparing
  e.g. Figs.~\ref{layers}(f,i). When the light ions reach the
  interface to vacuum, they are accelerated by the strong sheath-field
  there.

%Initially, most protons have low speeds at the transition to vacuum
%(as they lose energy as they cross the potential barrier), but they
%are accelerated efficiently by the quasi-static sheath field on the
%rear side.

%that the energy spectrum of light ions inside the heavy ion part
%of the target is broader than after the second layer of the target,
%ranging from zero for ions that had initial energy just above the
%threshold for reflection to the highest energy of reflected ions
%reduced by the size of the potential barrier at the interface. The
%reason for why the energy spectrum is broader is that partially
%reflected ions with sufficiently high energies continue through the
%heavy ion layer, typically at small speeds, as the shock hits the
%interface. As mentioned before, the number of such ions can to a large
%extent be minimized by appropriate choice of the relative density of
%the layers. 
}

\begin{figure}[htbp]
  \includegraphics[width=\columnwidth]{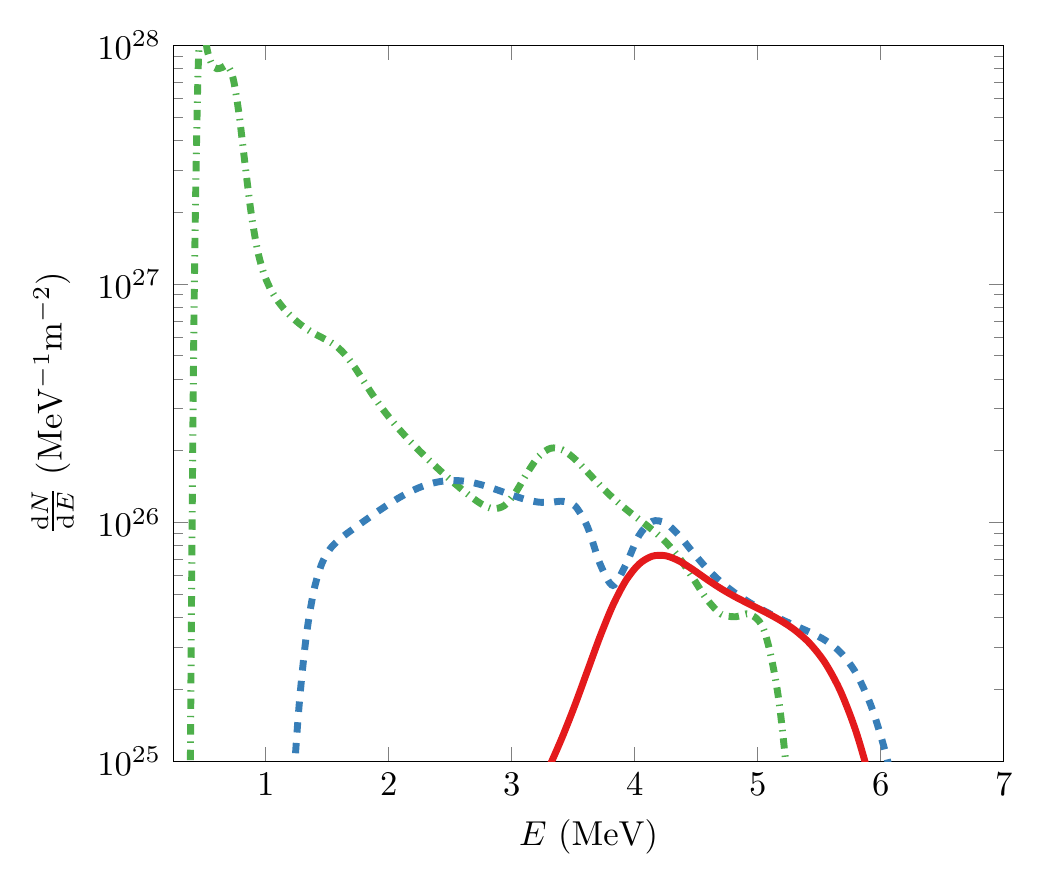}
  \caption{\label{protonslayers} Proton spectrum {at
     $t=108T$} for single-species and double-layered targets. Green
    dash-dotted line is for the single-species target.  Blue dashed
    line is for the layered target with $n=2.5 n_c$. Red solid line is
    for the layered target with the high density heavy ion layer
    having $n=25 n_c$. }
\end{figure}

In all cases the maximum proton energies exceed the energy of 2.9 MeV
for reflected ions by the shock-wave, as can be seen in
Fig.~\ref{protonslayers}, where the proton spectrums in the three
cases are presented. Furthermore, in the single species case we have a
broad TNSA-dominated proton spectrum. For the layered targets we
observe that the range of the spectrum shrinks and the maximum proton
energy increases compared to the single species case.  The shrinkage
of the spectrum is stronger in the high density
case. {In other words, by choosing the density of the
  heavy ion layer appropriately it should be possible to further
  optimize the monoenergeticity of the ion beam.} {As
  mentioned before, } the reason is that the longitudinal electric
field in the boundary region between the light and heavy ion part of
the layered target is stronger in the high density case, which hinders
the penetration of low energy ions to the high density
region. However, those ions that cross that boundary and reach the
rear side of the target will be efficiently accelerated.

The number of accelerated ions can be increased by using a thicker
proton layer on the front side of a double-layer target.  Then the
shock will be sustained for a longer time, as in the single species
case, where the shock is sustained throughout the whole target
width. To quantify the increase in particle number, we can compare the
number of shock-accelerated ions at different time instances in a
simulation with a target that has larger spatial extent. For example,
for the case with the exponential density profile and laser parameters
$a_0=2.5\sqrt{2}$, pulse length 25 fs, the numbers of shock
accelerated ions at $t=200T$ and $t=240T$ are 1.60 and 2.14 times
larger than that at $t=160T$.

\begin{figure}[htbp]
  \includegraphics[width=\columnwidth]{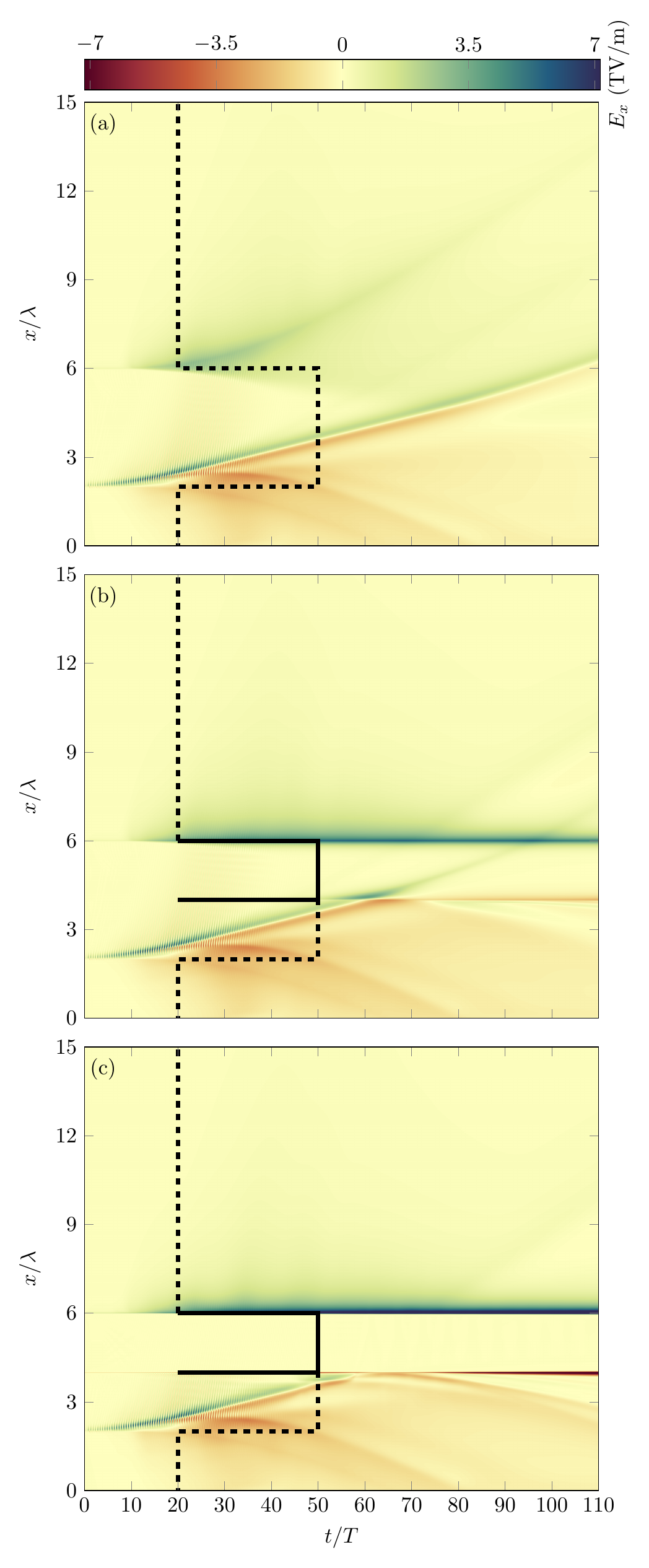}
  \caption{\label{Efield_layers} Longitudinal electric field as a
    function of position and time for targets irradiated by a linearly
    polarized pulse with $a_0= 2.5$ and pulse length
    \unit[50]{fs}. (a) shows the single-species target, (b) the
    layered target with $n=2.5n_c$ and (c) the layered target with the
    high density heavy ion layer having $n=25n_c$.  }
\end{figure}

Figure ~\ref{Efield_layers} shows the electric field as a function of
position and time in the layered targets compared with the single
species one. During the initial part of the simulation, the
sheath-field is set up by the hot electrons that are generated by the
laser-pulse. For the single-species target, the sheath field changes
its structure in time as the plasma expands at the rear boundary. On
the other hand, for the double layered targets, the sheath field is
stronger and has less time variation. The strongest sheath field is
obtained in the high density case, with a maximum value of 10
TV/m. Note that in the Fig.~\ref{Efield_layers} we only show values up
to the maximum value for the low density case (7 TV/m). {  Our simulations show that as long as the $A/Z\gtrsim 10$ the
  temporal variation of the sheath field will not affect the
  quality of the shock-accelerated protons.}

\section{Conclusions}
Vlasov-modelling of collisionless shock-acceleration allows for high
resolution of the distribution function, and therefore is highly
suitable in cases where effects of low-density tails in the
distribution function need to be resolved accurately.  In this paper
we have restricted the discussion to 1D1P modelling for
simplicity. Although the shock-dynamics is expected to be slightly
different in the 2D case, the main conclusions should be valid, given
the fact that particle-in-cell simulations have shown only a few
percent differences in the energy cutoff of the ions between 1D and 2D
configurations \cite{silva,lecz}.

We show that by using a target with a smooth (e.g.~exponentially
decreasing) density profile at the rear side, TNSA can be kept at a
low level, making CSA the main mechanism of acceleration of particles.
On the other hand, the energy of the shock-wave accelerated ions could
potentially be increased by the sheath-field produced at the rear
side. Provided that the sheath-field has limited variation in time,
mono-energeticity of the ions may be preserved.  Early launch of the
shock increases the number of ions that are reflected and can be
optimized by an appropriate choice of laser-parameters and also
potentially the density profile.

We observe that for the same laser fluency, a higher intensity
combined with a shorter laser pulse duration leads to a higher
shock-velocity, but the Mach-number is only slightly increased. The
main difference compared to the lower intensity and longer pulse
length case is that the shock develops later. This may be due that the
shorter duration of the laser pulse leads to a less peaked ion-density
and hence wider potential barrier, which results in a longer
reflection time for the ions. We show that splitting the laser-pulse
can also lead to higher shock-velocities, but without the delay of the
shock-formation.

Our simulations show that by using a target which consists of light
ions on the front side and heavy ions on the rear side, it is possible
to combine a strong quasi-static sheath-field with CSA. The dynamics
of shock-formation in the double-layer target resembles the one in a
rectangular single-species plasma slab, but as soon as the light ions
pass the rear side of the target, they obtain higher energies due to a
strong sheath field, which is produced due to the charge-separation
between the electrons that penetrate the rear side of the target and
the heavy ions at rest.  This leads to very efficient acceleration and
an increase in the proton energies compared to the energies of
shock-reflected ions, without broadening of the energy spectrum if the
heavy ion layer has high density.

\section{Appendix}

\subsection{Numerical description of one dimensional conservation laws}
Consider a conservation law:
\begin{equation}
\partial_t f+\partial_\zeta\left[a(\zeta,t) f\right]=0,\label{adv}
\end{equation}
which may be either Eq.~\eqref{tss1a} or Eq.~\eqref{tss2b}. Introduce the mapping $X(s,t,\zeta)$:
\begin{equation}
\left\{\begin{array}{l}
\frac{\text{d}X(s,t,\zeta)}{\text{d}s}=a[X(s,t,\zeta),s]\\
X(t,t,\zeta)=\zeta
\end{array}  \right.
\end{equation}
and a discretization $(\zeta_i,t_j)=(i\Delta \zeta,j\Delta t)$. It then holds that:
\begin{equation}
\int_{\zeta_{i-\frac{1}{2}}}^{\zeta_{i+\frac{1}{2}}}f(\zeta,t_{j+1})\,\text{d}\zeta=\int_{X(t_j,t_{j+1},\zeta_{i-\frac{1}{2}})}^{X(t_j,t_{j+1},\zeta_{i+\frac{1}{2}})}f(\zeta,t_{j})\,\text{d}\zeta,
\end{equation}
which can be written as:
\begin{equation}
\begin{array}{r}
\int_{\zeta_{i-\frac{1}{2}}}^{\zeta_{i+\frac{1}{2}}}f(\zeta,t_{j+1})\,\text{d}\zeta=\int_{\zeta_{i-\frac{1}{2}}}^{\zeta_{i+\frac{1}{2}}}f(\zeta,t_{j})\,\text{d}\zeta\\-\int_{X(t_j,t_{j+1},\zeta_{i+\frac{1}{2}})}^{\zeta_{i+\frac{1}{2}}}f(\zeta,t_{j})\,\text{d}\zeta\\+\int_{X(t_j,t_{j+1},\zeta_{i-\frac{1}{2}})}^{\zeta_{i-\frac{1}{2}}}f(\zeta,t_{j})\,\text{d}\zeta.
\end{array}\label{timeAdvancement}
\end{equation}
Introducing cell-averaged discrete values of the distribution function:
\begin{equation}
f_{i}^j=\frac{1}{\Delta \zeta}\int_{\zeta_{i-\frac{1}{2}}}^{\zeta_{i+\frac{1}{2}}}f(\zeta,t_{j})\,\text{d}\zeta
\end{equation}
 and fluxes $\phi_{i+\frac{1}{2}}$:
 \begin{equation}
\phi_{i+\frac{1}{2}}=\frac{1}{\Delta \zeta}\int_{X(t_j,t_{j+1},\zeta_{i+\frac{1}{2}})}^{\zeta_{i+\frac{1}{2}}}f(\zeta,t_{j})\,\text{d}\zeta,
\end{equation}
Eq.~\eqref{timeAdvancement} can be written as:
\begin{equation}
f_i^{j+1}=f_i^{j}-\phi_{i+\frac{1}{2}}+\phi_{i-\frac{1}{2}}.
\end{equation}
The choice of method to evaluate
$X(t_j,t_{j+1},\zeta_{i+\frac{1}{2}})$ and the corresponding flux
$\phi_{i+\frac{1}{2}}$ determines the accuracy of the method.

For the non-relativistic Vlasov-Poisson equation, $a(\zeta,t)$ is independent of $\zeta$ and
$X(t_j,t_{j+1},\zeta_{i+\frac{1}{2}})$ can be  determined to second order accuracy by:
\begin{equation}
X(t_j,t_{j+1},\zeta_{i+\frac{1}{2}})=\zeta_{i+\frac{1}{2}}-a(t_{j+\frac{1}{2}})\Delta t.\label{XCalc}
\end{equation}
For the relativistic Vlasov-Maxwell system on the other hand,
$a(\zeta,t)$ is not independent of $\zeta$  and Eq.~\eqref{XCalc}   yields a first
order accurate approximation of
$X(t_j,t_{j+1},\zeta_{i+\frac{1}{2}})$.

To evaluate the fluxes $\phi_{i+\frac{1}{2}}$, we use the positive and
flux conservative method \cite{filbet}.  The distribution function
$f(\zeta)$, in the cell with index $i$, is approximated in terms of the cell
averaged values with indices $(i-1)$, $i$ and $(i+1)$, according to:
\begin{align*}
f(\zeta)&=f_i+\frac{\epsilon_i^+}{6\Delta \zeta^2}[
2(\zeta-\zeta_i)(\zeta-\zeta_{i-\frac{3}{2}})\\
&+(\zeta-\zeta_{i-\frac{1}{2}})(\zeta-\zeta_{i+\frac{1}{2}})](f_{i+1}-f_{i})\\
&+\frac{\epsilon_i^-}{6\Delta \zeta^2}[ 2(\zeta-\zeta_i)(\zeta-\zeta_{i+\frac{3}{2}})\\
&+(\zeta-\zeta_{i-\frac{1}{2}})(\zeta-\zeta_{i+\frac{1}{2}})](f_{i}-f_{i-1}),
\end{align*}
where we have suppressed the time-index $j$. Furthermore, {the limiters} $\epsilon^+_i$ and $\epsilon^-_i$ are given by:
\begin{equation}
\epsilon^+_i=\left\{\begin{array}{lr}
\min\left(1,\frac{2f_i}{f_{i+1}-f_i}\right)&\text{ if
}f_{i+1}>f_i\\ \min\left(1,-2\frac{f_\infty-f_i}{f_{i+1}-f_{i}}\right)&\text{
  if }f_{i}>f_{i+1}
\end{array}\right.
\end{equation}
and:
\begin{equation}
\epsilon^-_i=\left\{\begin{array}{lr}
\min\left(1,2\frac{f_\infty-f_i}{f_{i}-f_{i-1}}\right)&\text{ if
}f_{i}>f_{i-1}\\ \min\left(1,\frac{-2f_i}{f_{i}-f_{i-1}}\right)&\text{
  if }f_{i-1}>f_{i}
\end{array}\right. .
\end{equation}
The quantity $f_\infty$ is the maximum cell-averaged
value. Straightforward integration yields the flux:
\begin{equation}
\begin{array}{r}
\phi_{i+\frac{1}{2}}=\alpha[f_{i}+\frac{\epsilon^+_{i}}{6}(1-\alpha)(2-\alpha)(f_{i+1}-f_{i})\\+\frac{\epsilon_{i}^-}{6}(1-\alpha)(1+\alpha)(f_{i}-f_{i-1})]\end{array}
\end{equation}
if $a_{i+\frac{1}{2}}$ is positive, where
$\alpha=[\zeta_{i+\frac{1}{2}}-X(t_j,t_{j+1},\zeta_{i+\frac{1}{2}})]/\Delta
\zeta$. For negative $a_{i+\frac{1}{2}}$, we instead have:
\begin{equation}\begin{array}{r}
\phi_{i+\frac{1}{2}}=\alpha[f_{i+1}-\frac{\epsilon^+_{i+1}}{6}(1-\alpha)(1+\alpha)(f_{i+2}-f_{i+1})\\-\frac{\epsilon_{i+1}^-}{6}(2+\alpha)(1+\alpha)(f_{i+1}-f_{i})].\end{array}
\end{equation}
 This is a third order interpolation of the fluxes, except in the
 presence of steep gradients. The limiters ensure that the
 interpolation is positivity preserving and does not violate the
 maximum principle. {Finally, as boundary conditions, we set the fluxes across boundaries to zero which enforces that particles can not leave or enter the domain and yields strict particle conservation.}

\subsection{Discretization of the electromagnetic field equations}
By introducing the quantities:
\begin{equation*}
G_{\pm}=E_z\pm cB_y\quad\text{and}\quad F_{\pm}=E_y\pm c B_z,
\end{equation*}
we may write:
\begin{eqnarray}
\left(\frac{\partial}{\partial t}\pm c\frac{\partial}{\partial
  x}\right)F_\pm=-J_y/\epsilon_0\label{b1}\\
\left(\frac{\partial}{\partial t}\pm c\frac{\partial}{\partial
  x}\right)G_\mp=-J_z/\epsilon_0.\label{EM}\label{b2}
\end{eqnarray}
Introducing characteristics $\eta=t+x/c$ and $\nu=t-x/c$, it holds  that:
\begin{eqnarray}
\left(\frac{\partial}{\partial t}+ c\frac{\partial}{\partial
  x}\right)F_{+}=2\frac{\partial F_+}{\partial
  \eta}\nonumber\\ \left(\frac{\partial}{\partial t}-
c\frac{\partial}{\partial x}\right)F_{-}=2\frac{\partial F_-}{\partial
  \nu}
\end{eqnarray}
as well as:
\begin{eqnarray}
\left(\frac{\partial}{\partial t}+ c\frac{\partial}{\partial x}\right)G_{-}=2\frac{\partial G_-}{\partial \eta} \nonumber\\\left(\frac{\partial}{\partial t}- c\frac{\partial}{\partial x}\right)G_{+}=2\frac{\partial G_+}{\partial \nu}.
\end{eqnarray}

To advance the equations \eqref{b1} and \eqref{b2}, we take $c\Delta
t=|\Delta x|$ and use a second order accurate central difference scheme:
\begin{equation}
F^{j+\frac{1}{2}}_{\pm,(i+\frac{1}{2}\pm 1)}=F_{\pm,(i+\frac{1}{2})}^{j-\frac{1}{2}}-\Delta t J^{j}_{y,(i+\frac{1}{2}\pm\frac{1}{2})}/\epsilon_0,
\end{equation}
\begin{equation}
G^{j+\frac{1}{2}}_{\pm,(i+\frac{1}{2}\mp 1)}=G_{\pm,(i+\frac{1}{2})}^{j-\frac{1}{2}}-\Delta t J^{j}_{z,(i+\frac{1}{2}\mp\frac{1}{2})}/\epsilon_0
\end{equation}
where $i$ is an index for the spatial-coordinate and $j$ is an index
for the temporal-coordinate.

Additionally, the electric field component $E_x$ is calculated by:
\begin{equation}
E_{x,(i+\frac{1}{2})}^{j+\frac{1}{2}}=\rho_i^{j+\frac{1}{2}}\Delta x+E_{x,(i-\frac{1}{2})}^{j+\frac{1}{2}},
\end{equation}
which is second order accurate, provided that the charge density can
be determined with first order accuracy.

{Regarding boundary conditions, the laser pulse is implemented as a Dirichlet boundary condition for the transverse fields and we use open boundary conditions at the boundary that is not associated with the laser. For the electric field component $E_x$ we have the  Dirichlet boundary condition $E_x=0$ at the right boundary. }

Finally, defining the discretized vector-potential on the spatial
cell-faces, it can be  calculated with second order accuracy in time on
integer time-steps by using a central-difference approximation of the
time-derivative in $ \partial \g A_\perp/\partial t=-\g E_\perp$.

\section*{Acknowledgements}
The authors are grateful to  E Siminos, J
Magnusson, A Stahl, I Pusztai and the rest of the {\sc pliona} team
for fruitful discussions.  This work was supported by the Knut and
Alice Wallenberg Foundation and the European Research Council
(ERC-2014-CoG grant 647121). The simulations were performed on
resources at Chalmers Centre for Computational Science and Engineering
(C3SE) provided by the Swedish National Infrastructure for Computing
(SNIC).

\end{document}